\documentclass[10pt,a4paper]{article}
\usepackage{amssymb}

\usepackage{graphicx}

\makeatletter
\renewcommand \thesection {\@Roman\c@section}
\makeatother
\input{tcilatex}

\begin{document}

\title{Quantum Key Distribution over 67 km with a plug\&play system}
\author{D. Stucki, N. Gisin, O. Guinnard*, G. Ribordy*, H. Zbinden \\
GAP-Optique, University of Geneva,\\
rue de l'Ecole-de-M\'{e}decine 20, CH-1211 Geneva 4\\
\ *id Quantique SA, rue Cingria 10, CH-1205 Geneva\\
email: hugo.zbinden@physics.unige.ch}
\date{}
\maketitle

\begin{abstract}
We present a fibre-optical quantum key distribution system. It works at
1550nm and is based on the plug\&play setup. We tested the stability under
field conditions using aerial and terrestrial cables and performed a key
exchange over 67 km between Geneva and Lausanne.
\end{abstract}

\section{Introduction}

Quantum cryptography or, more exactly, quantum key distribution (QKD) is the
most advanced subject in the field of quantum information technologies.
Since the introduction of the BB84 protocol~by Bennett and Brassard in 1984 
\cite{BB84} and their first implementation~in 1992\cite{B92}, many
experiments have been performed by numerous groups (see e.g. \cite{Gisin2001}
for a review). However, to our knowledge, all experiments to date were
performed in laboratories or used laboratory equipment (e.g. liquid nitrogen
cooled the detectors) or needed frequent alignments (e.g. control of
polarisation or phase). In this paper, we present a turn-key, fibre-optic
QKD-prototype that fits into two 19'' boxes, one for Alice and one for Bob
(see Fig.1). We tested the stability of the auto-compensating plug\&play
system \cite{muller} over installed terrestrial and aerial cables. Keys were
exchanged over a distance of 67 km.

We start with a short introduction to the plug\&play auto-compensating
setup, before describing the features of the prototype. We then recall the
relevant parameters of a QKD system and shortly discuss some security
issues. Finally the results of the field tests are presented.

\section{Plug\&Play Prototype}

Let's recall the principle of the so called plug\&play auto-compensating
setup \cite{muller, greg2000b, ibm, arhus, anders}, where the key is encoded
in the phase between two pulses travelling from Bob to Alice and back (see
Fig. 2). A strong laser pulse (@1550 nm) emitted at Bob is separated at a
first 50/50 beamsplitter (BS). The two pulses impinge on the input ports of
a polarisation beamsplitter (PBS), after having travelled through a short
arm and a long arm, including a phase modulator (PM$_{B}$) and a 50 ns delay
line (DL), respectively. All fibres and optical elements at Bob are
polarisation maintaining. The linear polarisation is turned by 90 degrees in
the short arm, therefore the two pulses exit Bob's setup by the same port of
the PBS. The pulses travel down to Alice, are reflected on a Faraday-mirror,
attenuated and come back orthogonally polarized. In turn, both pulses now
take the other path at Bob and arrive at the same time at the BS where they
interfere. Then, they are detected either in D$_{1}$, or after passing
through the circulator (C) in D$_{2}$. Since the two pulses take the same
path, inside Bob in reversed order, this interferometer is auto-compensated.
To implement the BB84 protocol, Alice applies a phase shift of $0$ or $\pi $
and $\frac{\pi }{2}$\ or $\frac{3\pi }{2}$ on the second pulse with PM$_{A}$%
. Bob chooses the measurement basis by applying a $0$ or $\frac{\pi }{2}$
shift on the first pulse on its way back.

\bigskip 

The prototype is easy to use. The two boxes just have to be connected via an
optical fibre. They are exclusively driven by two computers via the USB
port. The two computers communicate via a ethernet/internet link. The system
monitors on-line the temperature of the detectors, radiators and cases. The
photon counters are Peltier-cooled, actively gated, InGaAs/InP APD's \cite
{stucki2001}. The darkcount noise of the detectors is measured during the
initialization (the darkcount probability p$_{dark\text{ }}$is $\approx $10$%
^{-5}$ per gate). Although the setup needs no optical alignment, the phases
and the detection gates must be applied at the right time. Therefore, the
system measures in a next step the length of the link (the operator has only
to estimate the line's length at more or less 5 km). The variable attenuator
(VA) at Alice is set to a low level and bright laser pulses are emitted by
Bob. The time delay between the triggering of the laser and a train of gates
of the detectors is scanned until the reflected pulses are detected. The
delays for the two 2.5 ns detection gates are adjusted, as well as the
timing for the 50 ns pulse applied on the phasemodulator PM$_{B\text{ }}$.
In the plug\&play scheme, where pulses travel back and forth, (Rayleigh)
backscattered light can considerably increase the noise. Therefore, the
laser is not continuously pulsed, but trains of pulses are sent, the length
of these trains corresponding to the length of the storage-line introduced
for this purpose behind the attenuator at Alice's\cite{greg2000b}.
Consequently, the backward propagating pulses do no longer cross bright
pulses in the fibre. For a storage line measuring approximately 10 km, a
pulse train contains 480 pulses at a frequency of 5 MHz. A 90\% coupler (BS$%
_{10/90}$) directs most of the incoming light pulses to a APD-detector
module (D$_{A}$). It generates the trigger signal used to synchronise
Alice's 20 Mhz clock with the one of Bob. This synchronized clock allows
Alice to apply a 50 ns pulse at the phasemodulator PM$_{A}$ exactly when the
second, weaker pulse passes. Only this second pulse contains phase
information and must be attenuated below the one-photon-per-pulse level.
Measuring the height of the incoming pulses with D$_{A}$ would allow to
adjust the attenuator in order to obtain the right average number of photons
per outgoing pulse. For this purpose, the attenuator and the detector must
be calibrated beforehand. In practice, we measure the incoming power with a
power-metre. Random numbers are generated on both sides with a quantum
random number generator \cite{stefanov}. At Bob, clicks from each of the
photon counters are written together with the index of the pulse into a
buffer and transferred to the computer.

As a measure of security, the number of coincident clicks at both detectors
is registered, which is important to limit beamsplitting attacks (see
below). Morover, the incoming power at Alice is continuously measured with D$%
_{A}$, in order to detect so called Trojan horse attacks.

\section{Key parameters in QKD \label{parameters}}

\subsection{Key and error rates}

The first important parameter is the raw key rate $R_{raw}$ between Alice,
the transmitter, and Bob, the receiver: 
\begin{equation}
R_{raw}=q\nu \mu t_{AB}t_{B}\eta _{B}
\end{equation}
where $q$ depends on the implementation ($\frac{1}{2}$ for BB84 protocol,
because half the time Alice and Bob bases are not compatible), $\nu $ is the
repetition frequency, $\mu $ is the average number of photons per pulse, $%
t_{AB}$ is the transmission on the line Alice-Bob, $t_{B}$ is Bob's internal
transmission ($t_{B}\approx 0.6$) and $\eta _{B}$ is Bob's detection
efficiency ($\eta _{B}\approx 0.1$).

After $R_{raw}$ the second most important parameter is the quantum bit error
rate (QBER) which consists of four major contributions: 
\begin{equation}
QBER=\frac{false\,counts}{total\,counts}%
=QBER_{opt}+QBER_{dark}+QBER_{after}+QBER_{stray}  \label{QBER}
\end{equation}
$QBER_{opt}$ is simply the probability for a photon to hit the false
detector. It can be measured with strong pulses, by always applying the same
phases and measuring the ratio of the count rates at the two detectors. This
is a measure of the quality of the optical alignment of the polarisation
maintaining components and the stability of the fibre link. In the ideal
case, $QBER_{opt}$ is independent of the fibre length. $QBER_{dark}$ and $%
QBER_{after}$, the errors due to darkcounts and afterpulses, depend on the
characteristics of the photon counters\cite{stucki2001}. $QBER_{dark}$ is
the most important, it is the probability to have a darkcount per gate $%
p_{dark}$, divided by the probability to have a click $p_{det}$. 
\[
QBER_{dark}\cong \frac{p_{dark}}{\mu t_{AB}t_{B}\eta _{B}} 
\]
$QBER_{dark}$ increases with the distance and consequently limits the range
of QKD. $QBER_{after}$ is the probability to have an after pulse $%
p_{after}(t)$ summed over all gates between to detections: 
\begin{equation}
QBER_{after}\cong \sum_{n=0}^{n=\frac{1}{p_{det}}}p_{after}\left( \tau +n%
\frac{1}{\nu }\right)
\end{equation}
$p_{after}$, depending on the type of APD and on the temperature, decreases
rapidly with time\cite{stucki2001}. Nevertheless for high pulse rates ($\nu $%
= 5 MHz) $QBER_{after}$ can become significant. For instance, for $%
p_{det}=0.15\%$ (corresponding to about 7 dB loss with $\mu =0.1)$ we
measured a $QBER_{after}$ of about 4\%. By introducing a dead time $\tau $
of 4 $\mu s$ (during this time, following a detection, no gates are applied)$%
,$ $QBER_{after}$ can be reduced to 1.5\%. The bit rate $R_{raw}$\ on the
contrary, is only slightly reduced by a factor $\eta _{\tau }$: 
\begin{equation}
\eta _{\tau }=\frac{1}{1+\nu p_{det}\tau }\lessapprox 1
\end{equation}
In this example, $\eta _{\tau }$ becomes 0.97 and 0.92, for 4 and 12 $\mu s,$
respectively. In our prototype the deadtime can be varied between 0 and 12 $%
\mu s.$ The optimum deadtime varies as a function of distance, in our
measurements, however, we applied a constant deadtime of 4 $\mu s.$ Finally, 
$QBER_{stray}$, the errors induced by stray-light, essentially Rayleigh
back-scattered light, is a problem proper to the plug\&play setup. It can be
almost completely removed with the help of Alice's storage line and by
sending trains of pulses as mentioned above. However, we have to introduce
another factor $\eta _{duty}$ that reduces our bit rate. It gives the duty
cycle of the emitted pulse-trains and depends on the length of Alice's delay
line $l_{D}$ and the length of the fibre link $l_{AB}$:

\begin{equation}
\eta _{duty}=\frac{l_{D}}{l_{AB}+l_{D}}
\end{equation}
Hence with our prototype we can expect a raw rate of $R_{raw}$ of about:

\begin{equation}
R_{raw}=q\nu \mu t_{AB}t_{B}\eta _{B}\eta _{duty}\eta _{\tau }\approx
140kHz\left( \mu t_{AB}\frac{l_{D}}{l_{AB}+l_{D}}\right)
\end{equation}

\subsection{Error correction, privacy amplification and eavesdropping}

The net secret key rate is further reduced during the error correction and
privacy amplification process by a factor of $\eta _{dist}$. We didn't
implement error correction and privacy amplification for our field tests,
but we would like to estimate roughly the net key rate that could be
obtained with our system. In theory, $\eta _{dist}$ is simply given as the
difference between the mutal information of Alice and Bob, $I_{AB}$, and
Alice and Eve, $I_{AE}$\cite{Gisin2001}: 
\begin{equation}
\eta _{dist}=I_{AB}(D)-I_{AE}  \label{e1}
\end{equation}
Due to the errors, $I_{AB}$ is smaller than 1. It is a function of the
disturbance $D$, which is equal to the total $QBER$: 
\begin{equation}
I_{AB}=1+Dlog_{2}D+(1-D)log_{2}(1-D)  \label{e2}
\end{equation}
In the following we estimate the information of Eve, $I_{AE}$. In the line
of Felix et.al. \cite{felix2001} we make the following assumptions:

- The measured QBER should, within the statistical limits, be equal to what
is estimated according eq.\ref{QBER}. If this is not the case, a real user
won't proceed and blindly apply privacy amplification, he will stop the key
exchange and look for the problem. If the QBER is within these limits, we
attribute to Eve the $QBER_{opt}$ ($\lesssim $0.5\%) plus the error (2$%
\sigma )$ of the error estimation ($\lesssim 0.5\%$ for reasonably long
keys), say 1\% in total. In the case of perfect equipment of the
eavesdropper and true single photon source this error corresponds to an
information of $\frac{2}{ln2}1\%\cong 3\%$\cite{fuchs}.

- In the case of faint laser pulses and especially in the presence of high
fibre losses, Eve can take advantage of multi-photon pulses and gain
information while creating less or no errors\cite{felix2001}. In this case,
it is important to measure the length of the line and to register coincident
clicks at Bob's two detectors in order to limit Eve's possibilities. We
assume that Eve possesses perfect technology, but cannot efficiently measure
the number of photons without disturbing them and cannot store them.
Further, she disposes of fibers with losses as low as 0.15dB/km. Under these
assumptions one can calculate Eve's information per bit due to multi-photon
pulses $I_{2\nu }$ and obtains about 0.06, 0.14 and 0.40 for, 5, 10 and 20
dB loss, respectively (for $\mu =0.2$ , 0.25dB/km fiber loss and 10$^{8}$
pulses sent). Consequently, we obtain 
\begin{equation}
I_{AE}\cong 0.03+I_{2\nu }  \label{e3}
\end{equation}
With equations \ref{e1},\ref{e2} and \ref{e3} we can caluclate a theoretical
value of $\eta _{dist}$. In practice, $\eta _{dist}$\ will be smaller due to
the limitations of the used algorithm. Privacy amplification can be
performed without additional bit loss in contrary to error correction. For
our estimation, we use the results of Tancewsky et al \cite{tan} for $%
I_{AB}^{\prime }$ after error correction 
\begin{equation}
I_{AB}^{\prime }=1+Dlog_{2}D-\frac{7}{2}D
\end{equation}
which is in fact considerably smaller than $I_{AB}.$ The information of Eve $%
I_{AE}$ is reduced by the same factor $\frac{I_{AB}^{\prime }}{I_{AB}}$,
too. Finally, we obtain the following estimation of $R_{net}$: 
\begin{eqnarray}
R_{net} &=&\eta _{dist}R_{raw}\cong (I_{AB}^{{}}-I_{AE})\frac{I_{AB}^{\prime
}}{I_{AB}}R_{raw}  \label{netrate} \\
&\approx &\left[ 1+Dlog_{2}D-\frac{7}{2}D-(0.03+I_{2\nu
})(1-(1-D)log_{2}(1-D)-\frac{7}{2}D)\right] R_{raw}
\end{eqnarray}

\section{\label{security}Field measurements\label{results}}

\subsection{Visibilities}

In principle the prototype can be tested in the lab by performing key
exchange with different fibre losses and compare the measured QBER and bit
rates with the estimated values according to the simple formulas developed
above. There are two motivations for field tests on installed cables. The
first reason is to check, if the auto-compensating setup is robust in many
different situations. Several effects could reduce the visibility of the
interference. First, we have previously shown that Faraday rotation due to
the earth magnetic field cannot considerably decrease the visibility \cite
{Hugo2000}. Second, the time delay between the two pulses, travelling back
and forth between Alice and Bob, could change due to a temperature drift.
Let's assume that the temperature of the fiber increases with a rate $\theta %
\left[ \frac{K}{h}\right] .$ The time delay $\Delta t$ between the two
pulses is 54ns. If $\theta $ is constant for the whole trip of the pulses,
the second pulse will see a fibre that is longer by $\Delta l:$%
\begin{equation}
\frac{\Delta l}{l}=\alpha \Delta T
\end{equation}
\begin{equation}
\Delta l=\alpha 2l_{AB}\Delta T=\alpha 2l_{AB}\theta \Delta t
\end{equation}
With $\alpha =10^{-5}\left[ \frac{1}{K}\right] ,$ $l_{AB}=50km,$ $\theta =10%
\left[ \frac{K}{h}\right] $ we obtain $150pm\ll \lambda .$ Hence this effect
should be negligable especially since installed fibres have slow temperature
drifts. On the contrary, slow temperature induced length drifts can be large
enough that frequent readjustment of Bob's delay become necessary. In fact,
we noticed that during the heating up of Alice's box within the first hour
of operation, the changes in the delay line require a recalibration every 10
minutes or so. However, a bad synchronisation of the detection window does
not affect $QBER_{opt}$. Finally, mechanical stress could change the fiber
length and/or birefringence. If the birefringence changes rapidly, the
pulses are no longer orthogonally polarized at the input of Bob, despite the
Faraday mirror. In this case the two pulses might suffer different losses at
Bob's polarizing beamsplitter and the interference will no longer be
perfect. Rapid changes in stress are unlikely in installed cables, a couple
of meters below the surface. For this reason we tested the prototype also
over an aerial cable. We had at our disposition two fibres of 4.35 km
length, whereof 2.5 km in an aerial cable. In order to amplify a
hypothetical effect we put Alice and Bob side by side and passed twice
through the cable (config. A). In configuration B we inserted 1 spool of
about 15 km at the other end of the cable. Hence, the pulses made the
following trip: Bob, the aerial cable, 15 km spool, the aerial cable, 15 km
Alice with her 10 km storage line and back.

To measure the visibilities we send relatively strong pulses (a couple of
photons per pulse) with always the same, compatible phase values and look at
the counts on the two detectors, $R_{right}$ and $R_{wrong}$ (substracting
the counts due to detector noise). We obtain then the fringe visibility
according to the standard definition: 
\begin{equation}
V=\frac{R_{right}-R_{wrong}}{R_{right}+R_{wrong}}
\end{equation}

and the corresponding $QBER_{opt}$: 
\begin{equation}
QBER_{opt}=\frac{1-V}{2}
\end{equation}
Table 1 summerises the result of visibility measurements over different
cables. The indicated visibilities are the mean values over all four
possible compatible phase settings. There was no considerable decrease of
the visibility in any fibre, hence the auto-compensating interferometers
worked well under all tested conditions.

\bigskip

$
\begin{tabular}{||c|c|c|c||}
\hline\hline
fibre & length [km] & loss [dB] & Visibility [\%] \\ \hline
Geneva-Nyon (under lake) & 22.0 & 4.8 & 99.70 $\pm $ 0.03 \\ \hline
Geneva-Nyon (terrestrial) & 22.6 & 7.4 & 99.81 $\pm $ 0.03 \\ \hline
Nyon-Lausanne (terrestrial) & 37.8 & 10.6 & 99.63 $\pm $ 0.05 \\ \hline
Geneva-Lausanne (under lake) A & 67.1 & 14.4 & 99.62 $\pm $ 0.06 \\ \hline
Geneva-Lausanne (under lake) B & 67.1 & 14.3 & 99.66 $\pm $ 0.05 \\ \hline
Ste Croix (aerial) A & 8.7 & 3.8 & 99.70 $\pm $ 0.01 \\ \hline
Ste Croix (aerial) B & 23.7 & 7.2 & 99.71 $\pm $ 0.01 \\ \hline\hline
\end{tabular}
$

Table 1: Visibility measurements on different fibres

\bigskip

We tried to simulate an extremely unstable fibre link in the lab. For this
purpose, we put a fibre-optical polarisation scrambler (GAP-optique) at the
output of Bob followed by 25 km of fibre. We measured the visibility as
function of the scrambler frequency. This frequency is defined as the number
of complete circles that the vector of polarisation would describe per
second on the Poincar\'{e} sphere, if the birefringence changed uniformly.
The visibility drops from 99.7\% to 99.5\% and 98\% at frequencies of 40 Hz
and 100 Hz respectively. This shows that the visibilities can decrease under
rapid perturbations, however, it's unlikely to find such conditions using
installed fibres.

\subsection{Key exchange}

We performed key exchange over different installed cables, the longest
connecting the cities of Lausanne and Geneva (see Fig.3). We used always the
same file of random numbers, in a way that Bob could make the sifting and
calculation of error rate without communication. We estimated the net key
rate using equation (\ref{netrate}). Table 2 gives an overview of the
exchanged keys with $\mu =0.2$.

\bigskip

\begin{tabular}{||c|c|c|c||c||c||}
\hline\hline
fibre & length [km] & Key [kbit] & R$_{raw}[kHz]$ & QBER [\%] & R$%
_{net}[kHz] $ \\ \hline
Geneva-Nyon (under lake) & 22.0 & 27.9 & 2.06 & 2.0$\pm 0.1$ & 1.51 \\ \hline
Geneva-Nyon (terrestrial) & 22.6 & 27.5 & 2.02 & 2.1$\pm 0.1$ & 1.39 \\ 
\hline
Nyon-Lausanne (terrestrial) & 37.8 & 25.1 & 0.50 & 3.9$\pm 0.2$ & 0.26 \\ 
\hline
Geneva-Lausanne (under lake) A & 67.1 & 12.9 & 0.15 & 6.1$\pm 0.4$ & 0.044
\\ \hline
Geneva-Lausanne (under lake) B & 67.1 & 12.9 & 0.16 & 5.6$\pm 0.3$ & 0.051
\\ \hline
Ste Croix (aerial) A & 8.7 & 63.8 & 6.29 & 3.0$\pm 0.1$ & 4.34 \\ \hline
Ste Croix (aerial) B & 23.7 & 117.6 & 2.32 & 3.0$\pm 0.1$ & 1.57 \\ 
\hline\hline
\end{tabular}

Table 2: Overview of exchanged keys over different fibres ($\mu =0.2)$.

\bigskip

We notice that secure key exchange is possible over more than 60 km with
about 50 Hz of net key rate.

\section*{Conclusion}

We presented a QKD prototype, which can be simply plugged into the wall,
connected to a standard optical fibre and a computer via the USB\ port. It
allows key exchange over more than 60 km, with a net key rate of about 60
bits per second. The system is commercially available \cite{idq}.

\section{Figures}

Figure 1: Picture of the p\&p system\bigskip 

Figure 2: Schematic of the p\&p prototype\bigskip 

Figure 3: Satellite view of Lake Geneva with the cities of Geneva, Nyon and
Lausanne.\bigskip 

\section{Acknowledgments}

We would like to thank Michel Peris and Christian Durussel from Swisscom for
giving us access to their fibre links, as well as Laurent Guinnard and Mario
Pasquali for their help with the software and firmware, Jean-Daniel Gautier
and Claudio Barreiro for their help with the electronics. Finally, we thank
R\'{e}gis Caloz for the satellite picture. This work was supported by the
Esprit project 28139 (EQCSPOT) through Swiss OFES and the NCCR ''Quantum
Photonics.


\begin{thebibliography}{99}
\bibitem{BB84}  Ch.H. Bennett, and G. Brassard, ''Quantum cryptography:
public key distribution and coin tossing'', Int. conf. Computers, Systems \&
Signal Processing, Bangalore, India, December 10-12, 175-179 (1984).

\bibitem{B92}  Ch.H. Bennett, F. Bessette, G. Brassard, L. Salvail, and J.
Smolin, ''Experimental Quantum Cryptography'', J. Cryptology \textbf{5},
3-28 (1992).

\bibitem{Gisin2001}  N. Gisin, G. Ribordy, W. Tittel, and H.Zbinden,
''Quantum Cryptograpy'', quant-ph/0101098, to be published in Rev. Mod.
Phys. (2002).

\bibitem{muller}  A. Muller, T. Herzog, B. Huttner, W. Tittel, H. Zbinden
and N.. Gisin, ''Plug\&play systems for quantum cryptography, Applied Phys.
Lett. \textbf{70}, 793-795 (1997).

\bibitem{greg2000b}  G. Ribordy, J.-D. Gautier, N. Gisin, O. Guinnard, H.
Zbinden, ''Fast and user-friendly quantum key distribution'', J. Mod. Opt., 
\textbf{47}, 517-531 (2000).

\bibitem{ibm}  D. Bethune and W. Risk, ''An auto-compensating fiber-optic
quantum cryptography system based on polarization splitting of light'', IEEE
J. Quantum Electron. \textbf{36}, 340-347 (2000).

\bibitem{arhus}  P.M. Nielsen, C. Schori, J.L. Sorensen, L. Savail, I.
Damgard and E. Polzik, Experimental quantum key distribution with proven
security against realistic attacks, J. Mod. Opt. \textbf{48} (13), 1921-1942
(2001).

\bibitem{anders}  M. Bourenane, D. Ljunggren, A. Karlsson, P. Jonsson, A.
Hening and J. P. Ciscar, Experimental long wavelength quantum cryptography:
fromsimgle-photon transmission to key extraction protocols, J. Mod. Opt. 
\textbf{47} (2/3), 563-579 (2000).

\bibitem{stucki2001}  D. Stucki, G. Ribordy, A. Stefanov, H. Zbinden, J.G.
Rarity, T. Wall, ''Photon counting for quantum key distribution with Peltier
cooled InGaAs APD's'', J. Mod. Opt. \textbf{48 }(13), 1967-1982 (2001).

\bibitem{stefanov}  A. Stefanov, O. Guinnard, L. Guinnard, H. Zbinden and N.
Gisin, ''Optical Quantum Random Number Generator'', J. Mod. Optics \textbf{47%
}, 595-598 (2000), commercially available from idQuantique,
www.idquantique.com.

\bibitem{felix2001}  S.F\'{e}lix, N. Gisin, A. Stefanov and H. Zbinden,
''Faint laser quantum key distribution: eavesdropping exploiting multiphoton
pulses'', J. Mod. Opt. \textbf{48}, 2009-2022 (2001).

\bibitem{tan}  L. Tancevski, B. Slutsky, R. Rao, S. Fainman, Proc. SPIE 
\textbf{3228}, 322 (1997).

\bibitem{fuchs}  C.A. Fuchs, N. Gisin, R.B. Griffiths, C.S. Niu, and A.
Peres, ''Optimal Eavesdropping in Quantum Cryptography I''. Phys. Rev. A 56,
1163-1172 (1997).

\bibitem{Hugo2000}  H. Zbinden, N. Gisin, B. Huttner, A. Muller, and W.
Tittel, ''Practical Aspects of Quantum Cryptographic Key Distribution'', J.
Cryptology \textbf{13}, 207-220 (2000).

\bibitem{idq}  id Quantique SA, www.idquantique.com
\end{thebibliography}
\end{document}